\documentclass[11pt]{article}

\usepackage[latin1]{inputenc}

\usepackage{amstext,amsmath,amssymb,amsfonts,bbm,amsmath}
\usepackage{extarrows}
\usepackage{xcolor} 
\usepackage{hyperref}
\hypersetup{
    colorlinks=false,
    menubordercolor=red,
    linkbordercolor=blue
}
\usepackage{authblk}
\usepackage{caption} 
\usepackage{epsfig} 
\usepackage{color} 
\usepackage{graphicx} 
\usepackage{braket} 
\usepackage{slashed} 
\usepackage{dsfont} 
\usepackage{amsthm}  
\usepackage{multirow}

\usepackage{cite}

\usepackage{psfrag}
\usepackage{tikz} 
\usetikzlibrary{calc, external, intersections, arrows.meta, arrows, patterns, plotmarks, shapes.geometric, pgfplots.groupplots}
\usepackage{pgfplots}
\pgfplotsset{compat=1.9}
\usetikzlibrary{patterns}
\usetikzlibrary{calc}

\usepackage{float}  
\usepackage{subfig}
\usepackage[lmargin=60pt,rmargin=60pt,tmargin=60pt,bmargin=65pt]{geometry}

\allowdisplaybreaks[4]

\numberwithin{equation}{section}

\theoremstyle{remark}

\begin{document}

\title{\bf Note on surface defects in multiscalar critical models}

\author[1]{Andrii Anataichuk}
\author[2]{Sabine Harribey}

\affil[1]{\normalsize\it 
Taras Shevchenko National University of Kyiv, Kyiv, Ukraine
\authorcr \hfill }

\affil[2]{\normalsize \it 
 School of Theoretical Physics, Dublin Institute For Advanced Studies, 10 Burlington Road, Dublin, Ireland
  \authorcr \hfill
  \authorcr
  emails: andanataychuk@gmail.com , sharribey@stp.dias.ie
 \authorcr \hfill }

\date{}

\maketitle

\hrule\bigskip

\begin{abstract}
This paper studies generic surface defects for multiscalar critical models using a perturbative $\epsilon$ expansion in $4-\epsilon$ dimensions. The beta functions of the defect couplings for a generic multiscalar bulk with quartic interactions are computed at first non-trivial order in $\epsilon$. Specific bulks of interest are then considered: $O(N)$, hypercubic, hypertetrahdral, and biconical $O(m)\times O(n)$. In each case, we compute fixed points for the defect couplings and determine the remaining bulk symmetry. Expanding beyond the $O(N)$ model, we find a greater variety of patterns of symmetry breaking.
\end{abstract}

\hrule\bigskip

\tableofcontents

\section{Introduction}
\label{sec:introduction}

Studying systems with defects or boundaries is interesting for a variety of reasons and has a long history. Introducing defects to conformal field theories (CFTs) gives rise to a new type of CFTs, called defect CFTs (dCFTs) that realise conformal symmetry in a smaller dimension, given by that of the defect \cite{McAvity:1995zd,Billo:2016cpy}. More precisely, if one starts with a $d$-dimensional CFT and introduces a $p$-dimensional defect, the conformal group $SO(d+1,1)$ will be broken down to the subgroup $SO(p+1,1)\times SO(d-p)$. Such systems provide a new framework to investigate properties of quantum field theories (QFT).
From a more practical point of view, defects can describe numerous physical situations, such as impurities or localised perturbations. For experimental realisations of such systems, see for example \cite{sigl1986order, mailander1990near, burandt1993near, alvarado1982surface, PhysRevA.19.866, PhysRevB.40.4696, PhysRevB.58.12038}.

A model of particular interest, and one of the most thoroughly studied classes of CFTs, is the $O(N)$ model. Although it is not exactly solvable, numerous approximation methods can be used to study it, such as the $\epsilon$ expansion\cite{Wilson:1971dc}, the $1/N$ expansion \cite{Vasiliev:2003ev, Moshe:2003xn}, and the conformal bootstrap \cite{Polyakov:1974gs, Kos:2013tga, Poland:2018epd}. Going beyond the $O(N)$ model, the $\epsilon$ expansion can be used to study a wide range of CFTs that are reached as endpoints of renormalisation group (RG) flows triggered by operators that break the $O(N)$ symmetry. Then, it is possible to investigate RG fixed points for various global symmetries by generalising to quartic multiscalar models \cite{Pelissetto:2000ek, Osborn:2017ucf, Rychkov:2018vya, Osborn:2020cnf}.
The next logical step is to study RG flows and CFTs that emerge when these fixed points are deformed by defects. 

In the case of the three-dimensional $O(N)$ model, research on defects and boundaries dates back many years \cite{AJBray_1977, Ohno:1983lma, gompper1985conformal, McAvity:1995zd, Diehl:1996kd} and recent renewed interest has uncovered new universality classes for surface defects \cite{Metlitski:2020cqy, Padayasi:2021sik, Toldin:2021kun, Krishnan:2023cff}.

For multiscalar models, introducing defects can lead to various fixed points breaking the bulk symmetry in many ways. The case of a line defect perturbing a quartic multiscalar model was studied in \cite{Pannell:2023pwz}. For a $O(N)$ bulk, a line defect can only break the bulk symmetry to $O(N-1)$, while for more complicated bulk symmetries, such as hypercubic, different patterns of symmetry breaking emerged. This also has experimental applications beyond theoretical interest. For example, in dimension three, the Heisenberg and hypercubic models are almost indistinguishable due to nearly identical critical exponents \cite{Pelissetto:2000ek}. As a line defect would break the bulk symmetries in different ways, these two models could then be distinguished without having to measure critical exponents. 
The case of an interface (defect of co-dimension one) was thoroughly studied in \cite{Harribey:2023xyv, Harribey:2024gjn}, and such defects can break the bulk symmetry in many ways, leading to a vast space of interface CFTs. More recently, defects with continuously adjustable dimension $p=2+\delta$ were considered in \cite{deSabbata:2024xwn}, while line and surface defects for the long-range $O(N)$ model were considered in \cite{Bianchi:2024eqm}.
The case of a surface defect for the $O(N)$ model was studied in \cite{Giombi:2023dqs, Trepanier:2023tvb, Raviv-Moshe:2023yvq}. It was found that the bulk symmetry is broken to $O(l)\times O(N-l)$ with $l\leq N$ at leading order in $\epsilon$. Other bulk symmetries were considered in \cite{Pannell:2024hbu} for symmetry-preserving surface defects.
In this paper, we will go beyond the $O(N)$ model and study a surface defect for a generic multiscalar model, generalising the work of \cite{Giombi:2023dqs, Trepanier:2023tvb, Raviv-Moshe:2023yvq,Pannell:2024hbu}. Our defect will be on the $(x_1,x_2)$ plane (with other coordinates set to zero) and will take the general form
\begin{equation}
    S_{defect}=\int  dx_1 dx_2 \frac{h_{ij}}{2}\phi_i\phi_j(x_1,x_2,\Vec{0}) \, , \qquad i,j=1,\dots, N \, , 
\end{equation}
where $\phi_i$ are scalar fields and $h_{ij}$ are defect couplings. This surface defect will be added to $(4-\epsilon)$-dimensional bulk CFTs with different global symmetry groups tuned to criticality. We will then compute fixed points for different values of $N$ and study the possible symmetry-breaking patterns. We will see that, depending on the bulk symmetry, we can have a larger variety of symmetry-breaking patterns than for the $O(N)$ bulk. 

The rest of this paper is organised as follows. In section \ref{sec:model}, we present the model in more detail and compute the defect beta functions at one loop for a generic multiscalar bulk using minimal subtraction. In section~\ref{sec:fixedpoints}, we classify surface defect fixed points. We start by using group theory methods to simplify the beta functions and derive properties of fixed points for a generic bulk. We then study various global bulk symmetries in more detail: free, $O(N)$, hypercubic, hypertetrahedral, and biconical $O(m)\times O(n)$ bulks. In each case, we compute fixed points analytically, determine the remaining symmetry, and study stability at the fixed points. 

\section{Model and beta functions}
\label{sec:model}

We will study a multiscalar model with quartic interactions in a $d$-dimensional bulk and quadratic defects localised on a surface. It is a generalisation of the models studied in \cite{Trepanier:2023tvb,Giombi:2023dqs,Raviv-Moshe:2023yvq} and is given by the following action
\begin{align}
        S &= \int d^d x \biggl [ \frac{1}{2} \partial_{\mu} \phi_i \partial^{\mu} \phi_i(x) + \frac{1}{4!} \lambda_{ijkl} \phi_i \phi_j \phi_k \phi_l(x) \biggl]  + \frac{h_{ij}}{2} \int dx_1 dx_2 \, \phi_i \phi_j(x_1,x_2,\Vec{0}) \, ,
    \label{eq:action}
\end{align}
where the indices take values from $1$ to $N$, and a summation over repeated indices is implicit. The couplings $\lambda_{ijkl}$ and $h_{ij}$ are fully symmetric thus corresponding in general to $\binom{N+3}{4}$ and $\binom{N+1}{2}$ couplings, respectively. In dimension four, both the bulk and surface interactions are marginal. In the following, we will thus set $d=4-\epsilon$ to allow for a perturbative treatment.
Recall that the propagator of the free theory is given by
\begin{equation}
    \langle \phi_i(x)\phi_j(y) \rangle =\delta_{ij} \int \frac{d^d p}{\left( 2\pi \right)^d}\frac{e^{ip \cdot (x-y)}}{p^2}=\delta_{ij}\frac{C_{\phi}}{|x-y|^{d-2}} \, , \qquad C_{\phi}=\frac{\Gamma\left( \tfrac{d-2}{2}\right)}{4\pi^{\tfrac{d}{2}}} \, .
\end{equation}

The theory in the bulk is not modified by the surface defect, and the beta functions for the bulk couplings are given at one loop by \cite{ZinnJustin:2002ru}
\begin{equation}
    \beta_{ijkl}=-\epsilon \lambda_{ijkl}+\left( \lambda_{ijmn}\lambda_{mnkl} + \lambda_{ikmn}\lambda_{mnjl} +\lambda_{ilmn}\lambda_{mnkm}\right) \, ,
\end{equation}
where the bulk couplings have been rescaled as $\lambda_{ijkl} \rightarrow (4\pi)^2 \lambda_{ijkl}$.

Following the method of \cite{Giombi:2023dqs}, we will compute the beta functions of the defect couplings to lowest order by requiring finiteness of the one-point function of $\phi_i\phi_j(0,x)$ at a distance $|x|$ from the defect. 

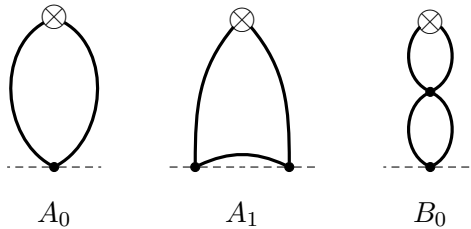
\begin{figure}[h!]
        \begin{tikzpicture}[baseline=(vert_cent.base), square/.style={regular polygon,regular polygon sides=4},scale=1.25]
        \node at (0,1.6) []  (t) {};
        \node at (0,0) [circle,draw,fill=black,inner sep=1.2pt,outer sep=0pt] (b) {};
        \draw[very thick] (t) to [out=-150,in=150] (b);
        \draw[very thick] (t) to [out=-30,in=30] (b);
        \draw[densely dashed] (b)--++(0:0.5cm);
        \draw[densely dashed] (b)--++(180:0.5cm);
        \node[inner sep=0pt,outer sep=0pt, scale=1.27] (vert_cent) at (0,1.6) {$\otimes$};
        \node at (0, -0.5) {$A_0$};

        \node at (2,1.6) []  (t) {};
        \node at (1.5,0) [circle,draw,fill=black,inner sep=1.2pt,outer sep=0pt] (bl) {};
        \node at (2.5,0) [circle,draw,fill=black,inner sep=1.2pt,outer sep=0pt] (br) {};
        \draw[very thick] (t) to [out=-130,in=90] (bl);
        \draw[very thick] (t) to [out=-50,in=90] (br);
        \draw[very thick] (bl) to [out=25,in=155] (br);
        \draw[densely dashed] (bl)--++(0:0.5cm);
        \draw[densely dashed] (bl)--++(180:0.3cm);
        \draw[densely dashed] (br)--++(0:0.3cm);
        \draw[densely dashed] (br)--++(180:0.5cm);
        \node[inner sep=0pt,outer sep=0pt, scale=1.27] (vert_cent) at (2,1.57) {$\otimes$};
         \node at (2, -0.5) {$A_1$};

        \node at (4,1.6) []  (t) {};
        \node at (4,0.8) [circle,draw,fill=black,inner sep=1.2pt,outer sep=0pt] (m) {};
        \node at (4,0) [circle,draw,fill=black,inner sep=1.2pt,outer sep=0pt] (b) {};
        \draw[very thick] (t) to [out=-135,in=150] (m);
        \draw[very thick] (t) to [out=-45,in=30] (m);
        \draw[very thick] (m) to [out=-150,in=150] (b);
        \draw[very thick] (m) to [out=-30,in=30] (b);
        \draw[densely dashed] (b)--++(0:0.5cm);
        \draw[densely dashed] (b)--++(180:0.5cm);
        \node[inner sep=0pt,outer sep=0pt, scale=1.27] (vert_cent) at (4,1.55) {$\otimes$};
         \node at (4, -0.5) {$B_0$};
    \end{tikzpicture}
    \centering
    \caption{Diagrams $A_0$, $A_1$, $B_0$ that contribute to the one-point function $\langle \phi_i \phi_j(0,x)\rangle$ at lowest order. The crossed circle represents the operator $\phi_i \phi_j$ while the black vertices represent insertion of bulk and defect couplings. The surface where the defect is localised is represented by the dashed line. Black vertices on the dashed line are thus defect couplings.}
    \label{fig:two_loops}
\end{figure}

To lowest order, there are three graphs contributing to the one-point function $\langle\phi_i\phi_j(0,x)\rangle$. They are represented in Fig.~\ref{fig:two_loops}. 

The renormalised one-point function $\langle [\phi_i\phi_j](0,x)\rangle$ is then given by
\begin{equation}
    \langle [\phi_i\phi_j](0,x)\rangle=\frac{1}{Z_{ijkl}}\left(h_{kl}A_0+ h_{km}h_{ml} A_1+\lambda_{klmn}h_{mn}B_0 \right) \, ,
    \label{eq:onepoint}
\end{equation}
with $Z_{ijkl}$ the wave function renormalisation factor
\begin{equation}
    \phi_i\phi_j(0,x)=Z_{ijkl}[\phi_k\phi_l](x)\, , \quad Z_{ijkl}=\frac{1}{2}\left( \delta_{ik}\delta_{jl}+\delta_{il}\delta_{jk}\right)-\frac{\lambda_{ijkl}}{(4\pi)^2 \epsilon} \, ,
\end{equation}
which is the same as for the bulk theory without defect \cite{Kleinert:2001ax}. 

The amplitudes $A_0$, $A_1$ and $B_0$ have been computed in \cite{Trepanier:2023tvb,Giombi:2023dqs} and are given by
\begin{align}
    A_0&=-\frac{1}{16 \pi^3 x^2} +\mathcal{O}(\epsilon) \, , \\
    A_1 &= \frac{1}{32 \pi^4 x^2 \epsilon}+ \mathcal{O}(1) \, ,  \\
    B_0 &= \frac{1}{(4\pi)^2 8 \pi^3 x^2 \epsilon}+\mathcal{O}(1) \, .
\end{align}
Following the method of \cite{Giombi:2023dqs}, we now substitute the bare couplings by the renormalised ones as
\begin{align}
    \lambda_{ijkl,B}&=\mu^{\epsilon} \left(\lambda_{ijkl,R} +\mathcal{O}(\lambda^2) \right) \, , \\
    h_{ij,B}&= \mu^{\epsilon}\left( h_{ij,R} +\delta h_{ij} \right) \, ,
\end{align}
where $\mu$ is an arbitrary mass scale that renders the renormalised couplings dimensionless. In the following, we will omit the subscripts $B$ and $R$. 
Substituting in \eqref{eq:onepoint} and requiring that the renormalised one-point function $\langle [\phi_i \phi_j](x)\rangle$ must be finite we find
\begin{equation}
    \delta h_{ij}= \frac{1}{2\pi \epsilon}h_{ik}h_{kj}+\frac{1}{(4\pi)^2 \epsilon}\lambda_{ijkl}h_{kl} \, .
\end{equation}

Finally, defining $\beta_{ij}=\mu \partial_{\mu} h_{ij}$, we obtain
\begin{equation}
    \beta_{ij}=-\epsilon h_{ij} + h_{ik}h_{kj}+ \lambda_{ijkl}h_{kl} \, ,
    \label{eq:genericbeta}
\end{equation}
where we have rescaled the couplings as $\lambda_{ijkl} \rightarrow (4\pi)^2 \lambda_{ijkl}$ and $h_{ij} \rightarrow (2\pi) h_{ij}$. The renormalised one-point function of the operator $\phi_i \phi_j$ is then given by
\begin{equation}
    \langle [\phi_i \phi_j](0,x)\rangle =-\frac{h_{ij}}{8 \pi^2 x^2} \, ,
\end{equation}
where we have also rescaled the couplings as $h_{ij} \rightarrow (2\pi) h_{ij}$.

\section{Classification of fixed points}
\label{sec:fixedpoints}

Fixed points related by a $O(N)$ transformation are equivalent up to a redefinition of the fields. Therefore, we can diagonalise $h_{ij}$ without loss of generality. We call its eigenvalues $ h_1,\dots , h_N$. Moreover, following the decomposition of \cite{Osborn:2020cnf}, the bulk coupling can be written as
\begin{align}
    \lambda_{ijkl}&=  \frac{d_0}{3} \left( \delta_{ij}\delta_{kl}+\delta_{ik}\delta_{jl}+\delta_{il}\delta_{jk}\right) \crcr
    &+ \frac{1}{6}\left( \delta_{ij}d_{2,kl}+ \delta_{kl}d_{2,ij}+\delta_{ik}d_{2,jl}+\delta_{jl}d_{2,ik}+\delta_{il}d_{2,jk}+\delta_{jk}d_{2,il} \right)\crcr
    & +  d_{4,ijkl} \, ,
    \label{eq:decomp}
\end{align}
where $d_{2,ij}$ and $d_{4,ijkl}$ are symmetric traceless tensors. 

Plugging this into \eqref{eq:genericbeta}, the beta functions become
\begin{equation}
   \beta_i=h_i^2 + \biggl(\frac{2}{3}d_0+\frac{5}{6}d_{2,ii}-\epsilon \biggl)h_i+ \biggl(\frac{d_0}{3}+\frac{1}{6}d_{2,ii} \biggl )\mathrm{Tr}(h)+d_{4,iikk}h_k \, ,
   \label{eq:betadecomp}
\end{equation}
where the index $i$ is not summed.

In all generality, this system of $N$ equations is too complicated to solve, and we have to specify either a bulk or a defect symmetry. However, we can make the following remark. For bulks where $d_{2,ii}$ and $d_{4,iikk}$ are equal for all $i=1,\dots, N$, the beta functions \eqref{eq:betadecomp} will be the same for all $i$. This means that all $h_i$ satisfy the same quadratic equation and can thus take only two values. The fixed points will then be given by taking $l$ couplings equal to some value $h_l$ and $N-l$ couplings equal to some other value $h_{N-l}$. If we started with a bulk symmetry $G_N$, such fixed points will break it to $G_l \times G_{N-l}$. We will see below that this is the case for the $O(N)$ and hypercubic bulks. 

\subsection{$O(N)$-invariant surface defect}

While keeping the bulk generic, we can look at the simple case $h_i=h$ for all $i=1,\dots,N$.  The surface defect is then $O(N)$-invariant and does not break the bulk symmetry. 
Plugging this in \eqref{eq:betadecomp}, the fixed-point equation becomes
\begin{equation}
    h\biggl(h+\frac{(N+2)}{3}d_0-\epsilon \biggl)=0 \, ,
\end{equation}
which gives one non-trivial fixed point
\begin{equation}
   h=\epsilon-\frac{(N+2)}{3}d_0 \, .
   \label{eq:fpone}
\end{equation}

The critical exponent for the non-trivial fixed point is given by $\omega=\epsilon-\frac{(N+2)}{3}d_0$, while for the trivial fixed point, it is $\omega_0=\frac{(N+2)}{3}d_0-\epsilon$. Therefore, if $d_0<\frac{3}{N+2}\epsilon$, the non-trivial defect is stable, and the trivial defect is unstable, while for $d_0>\frac{3}{N+2}\epsilon$, the trivial defect is stable, and the non-trivial defect is unstable. When $d_0=\frac{3}{N+2}\epsilon$, only the trivial fixed point exists, and it is marginal.

We will see below that for a $O(N)$ bulk and a hypercubic bulk at one loop, we are in the case $d_0<\frac{3}{N+2}\epsilon$, where the non-trivial defect is stable. 

Finally, the renormalised one-point function of $\phi_i \phi_j$ is given by
\begin{equation}
    \langle [\phi_i \phi_j](0,x)\rangle= \frac{(N+2)d_0-3\epsilon}{24\pi^2 x^2}\delta_{ij} \, .
\end{equation}

\subsection{Free bulk}

We will now keep the defect generic and specify a bulk symmetry. Let us start with the simple case of a free bulk. The beta functions \eqref{eq:genericbeta} simplify to
\begin{equation}
    \beta_i=h_i(h_i-\epsilon) \, .
\end{equation}
The equations $\beta_i=0$ have $2^N$ solutions falling into $N+1$ equivalence classes, breaking the bulk symmetry in different ways. The first solution is the trivial defect $h_i=0$ for all $i=1,\dots, N$, which does not break the $O(N)$ bulk symmetry. The other $N$ equivalence classes are obtained by taking $0<l\leq N$ couplings non-zero and equal to $h_l=\epsilon$ while the remaining $N-l$ couplings are set to zero. This will break the $O(N)$ symmetry of the bulk to $O(l) \times O(N-l)$.

The stability matrix is diagonal and has one eigenvalue of multiplicity $l$, $\omega_l=\epsilon$, and one eigenvalue of multiplicity $N-l$, $\omega_{N-l}=-\epsilon$. Therefore, the only stable fixed point is the one where all surface couplings are non-zero, while the trivial surface defect is fully unstable.

\subsection{$O(N)$ bulk}

The case of an interacting $O(N)$ bulk was studied in \cite{Trepanier:2023tvb,Giombi:2023dqs,Raviv-Moshe:2023yvq}, but let us check here that we recover their result. 

The interacting $O(N)$ bulk corresponds to $\lambda_{ijkl}=\frac{\lambda}{3}\left( \delta_{ij}\delta_{kl}+\delta_{ik}\delta_{jl}+\delta_{il}\delta_{jk}\right)$. Therefore $d_0=\lambda$, $d_{2,ij}=0$ and $d_{4,ijkl}=0$. 

At the bulk fixed point we have at one loop $\lambda=\frac{3\epsilon}{N+8}$ and the fixed-point equations become
\begin{equation}
    0=h_i^2-\epsilon\frac{N+6}{N+8}h_i+\epsilon\frac{\mathrm{Tr}(h)}{N+8} \, ,
\end{equation}
which is equivalent to equation $(2.10)$ of \cite{Trepanier:2023tvb}.

These equations can only be solved by taking $l$ couplings equal to $h_l$ and $N-l$ couplings equal to $h_{N-l}$, breaking the $O(N)$ symmetry to $O(l)\times O(N-l)$\footnote{This can also be interpreted as the original matrix $h_{ij}$ of surface defects having one eigenvalue of multiplicity $l$ and one eigenvalue of multiplicity $N-l$.}. More precisely the fixed points are given by the pair $(h_{l+},h_{N-l,-})$ where 
\begin{equation}
    h_{l\pm}=\frac{N+3-l\pm\sqrt{9-l(N-l)}}{N+8}\epsilon + \mathcal{O}(\epsilon^2)\, .
\end{equation}
For $l=0,N$, we obtain the trivial solution and the solution with all couplings equal \eqref{eq:fpone} with $h=\frac{6\epsilon}{N+8}$, which is stable 
 as $d_0=\frac{3\epsilon}{N+8}<\frac{3\epsilon}{N+2}$.

\subsection{Hypercubic bulk}

The hypercubic model has symmetry $B_N=\mathbb{Z}_2^N \rtimes S_N$, where $S_N$ is the group of permutations of $N$ objects. It has two quadratic couplings given by
\begin{equation}
    \lambda_{ijkl}=\lambda\left( \delta_{ij}\delta_{kl}+\delta_{ik}\delta_{jl}+\delta_{il}\delta_{jk}\right)+g d_{ijkl} \, ,
\end{equation}
where $d_{ijkl}=\delta_{ijkl}-\frac{1}{N+2}\left( \delta_{ij}\delta_{kl}+\delta_{ik}\delta_{jl}+\delta_{il}\delta_{jk}\right)
$ and $\delta_{ijkl}$ is a generalisation of the Kronecker delta to four indices.

We therefore have $d_0=3\lambda$, $d_{2,ij}=0$ and $d_{4,ijkl}=g d_{ijkl}$. At the non-trivial fixed point $\lambda=2\epsilon\frac{N-1}{3N(N+2)}$ and $g=\epsilon\frac{N-4}{3N}$. Substituting into \eqref{eq:betadecomp}, we obtain
\begin{equation}
    \beta_i=h_i^2-2\epsilon\frac{N+1}{3N}h_i+\epsilon\frac{\mathrm{Tr}(h)}{3N} \, .
\end{equation}

These equations have the same form as for the $O(N)$ bulk and will be solved by taking $l$ couplings equal to $h_l$ and $N-l$ couplings equal to $h_{N-l}$. 

The fixed points are then given by solutions of the following system
\begin{align}
   0&= h^2_l - 2\epsilon \frac{N+1}{3N} h_l + \frac{\epsilon}{3 N} (l h_l + (N-l) h_{N-l})  \, ,\crcr
    0& =  h^2_{N-l} - 2 \epsilon\frac{N+1}{3N} h_{N-l} + \frac{\epsilon}{3 N} (l h_l + (N-l) h_{N-l}) \, .
\end{align}
At first order in $\epsilon$, there are two non-trivial solutions
\begin{align}
   h_{l,+}&=\frac{\epsilon}{6 N} \bigl(2+3N - 2l + \sqrt{(N+2)^2-4l(N-l)} \bigl)  \, ,
   \crcr
   h_{N-l,-}&=\frac{\epsilon}{6 N} \bigl(2+N + 2l - \sqrt{(N+2)^2-4l(N-l)} \bigl) \; ,
\label{eq:fpcub1}
\end{align}
and
\begin{align}
 h_{l,-}&=\frac{\epsilon}{6 N} \bigl(2+3N - 2l - \sqrt{(N+2)^2-4l(N-l)} \bigl) \, ,\crcr
 h_{N-l,+}&=\frac{\epsilon}{6 N} \bigl(2+N + 2l + \sqrt{(N+2)^2-4l(N-l)} \bigl)\; .
 \label{eq:fpcub2}
\end{align}

Notice that substituting $l \rightarrow N-l$ into the second solution gives the first solution. Thus, we can consider only the first solution. 

These fixed points break the bulk symmetry $B_N$ to $B_{l}\times B_{N-l}$. If $l=0$ or $l=N$, we recover the $O(N)$-invariant defect with the non-trivial fixed point located at $h=\epsilon\tfrac{N+2}{3N}$.

At the fixed point \eqref{eq:fpcub1}, the one-point function of $\phi_i^2$ is given by
\begin{align}
      \langle [\phi_l^2](0,x)\rangle & = \frac{2l-2-3N-\sqrt{(N+2)^2-4l(N-l)}}{48\pi x^2} \epsilon\, , \crcr
      \langle [\phi_{N-l}^2](0,x)\rangle & = \frac{-(2+N+2l)+\sqrt{(N+2)^2-4l(N-l)}}{48\pi x^2}\epsilon \, .
\end{align}

\paragraph{Stability matrix}

The critical exponents are given by the eigenvalues of the stability matrix $S_{ij}=\frac{\partial \beta_i}{\partial h_j}$, which can be expressed as
\begin{align}
    S_{ii}&=2 h_i-\frac{2N+1}{3N}\epsilon \, , \crcr
    S_{ij}&=\frac{\epsilon}{3N} \, , \, i\neq j \, .
\end{align}

Due to the symmetric nature of the stability matrix, it is possible to compute its eigenvalues for all $N$. 
For $l\neq 0,N$, we find two eigenvalues of multiplicity one, $\omega_{\pm}$, one eigenvalue of multiplicity $l-1$, $\omega_{l-1}$ and one eigenvalue of multiplicity $N-l-1$, $\omega_{N-l-1}$. If $l=1$, $\omega_{l-1}$ does not exist and if $l=N-1$, $\omega_{N-l-1}$ does not exist.

At the fixed point \eqref{eq:fpcub1} and at first order in $\epsilon$, the critical exponents are given by
\begin{align}
    \omega_{\pm}&=  \frac{\epsilon}{6N} \bigl ( N \pm \sqrt{N^2+4b(N-2l+b)} \bigl) \, , \crcr
    \omega_{l-1}&=-\omega_{N-l-1}=\frac{N-2l+b}{3N}\epsilon \, ,
\end{align}
where $b=\sqrt{(N+2)^2-4l(N-l)}$. 

We have $\omega_+>0$, $\omega_-<0$, $\omega_{l-1}>0$ and $\omega_{N-l-1}<0$. Therefore, we always have at least one negative critical exponent: the fixed points are unstable. 

Let us now look at the $O(N)$-invariant defect (obtained by setting $l=N$ in \eqref{eq:fpcub1}). 
In this case, we have an eigenvalue of multiplicity $N-1$, $\omega_{N-1}=\frac{2\epsilon}{3N}$ and an eigenvalue of multiplicity one, $\omega_1=\frac{N+2}{3N}\epsilon$. They are both positive: the fixed point is stable. 

Let us finally look at the trivial fixed point (obtained by setting $l=0$ in \eqref{eq:fpcub1}). We now have an eigenvalue of multiplicity $N-1$, $\omega_{N-1}=-\frac{2(N+1)\epsilon}{3N}$ and an eigenvalue of multiplicity one, $\omega_1=-\frac{N+2}{3N}\epsilon$. They are both negative: the trivial fixed point is unstable.

We conclude that the only stable fixed point is the one with all defect couplings equal to $h=\epsilon\tfrac{N+2}{3N}$, which does not break the hypercubic bulk symmetry. 

\subsection{Hypertetrahedral bulk}

We now consider a bulk with hypertetrahedral symmetry $T_N=S_{N+1}\times \mathbb{Z}_2$. To do so we introduce $N+1$ vectors $(e_N)_i^\alpha$, $\alpha=1,\ldots,N+1$,\footnote{These vectors point to the $N+1$ vertices of an $N$-dimensional hypertetrahedron.} satisfying
\begin{equation}
    \sum_{\alpha} (e_N)_i^\alpha=0\,,\qquad \sum_{\alpha}(e_N)^\alpha_i (e_N)_j^\alpha=\delta_{ij}\,,\qquad (e_N)_i^\alpha (e_N)_i^\beta=\delta^{\alpha\beta}-\frac{1}{N+1}\,.
    \label{eq:evectors}
\end{equation}
We use the conventions of \cite{Pannell:2023pwz} to define these vectors recursively. For $N=1$ we set
\begin{equation}
    (e_1)_1^1=-(e_1)_1^2=-\tfrac{1}{\sqrt{2}} \, ,
\end{equation}
and for $N>1$
\begin{align}
    (e_N)_i^{\alpha}&=(e_{N-1})_i^{\alpha} \, , \quad i=1,\dots, N-1 \, , \, \alpha=1,\dots , N \, , \crcr
    (e_N)_N^{\alpha}&=-\frac{1}{\sqrt{N(N+1)}} \, , \quad \alpha=1,\dots, N \, , \crcr
    (e_N)_i^{N+1}&=\sqrt{\frac{N}{N+1}}\delta_i^N \, .
\end{align}

The hypertetrahedral model is obtained by taking $d_0=3\lambda$, $d_{2,ii}=0$ and $d_{4,ijkl}=g d_{ijkl}$ with
\begin{equation}
    d_{ijkl}=\sum_{\alpha} (e_N)_i^{\alpha}(e_N)_j^{\alpha}(e_N)_k^{\alpha}(e_N)_l^{\alpha}-\frac{N}{(N+1)(N+2)}\left( \delta_{ij}\delta_{kl}+\delta_{ik}\delta_{jl}+\delta_{il}\delta_{jk}\right) \, .
\end{equation}

We have to be careful in our analysis of the beta functions. Due to the nature of the $e_N$ vectors, even starting with a diagonal matrix of defect couplings, the renormalisation flow will generate non-diagonal couplings. We decide to tune them to zero, but we still have to make sure the corresponding beta functions vanish at the fixed point.\footnote{We thank William Pannell for pointing this out.} 

The beta functions then become:
\begin{align}
    \beta_i=&h_i^2 +\left( 2\lambda-\frac{2N}{(N+1)(N+2)}g-\epsilon\right)h_i+\left( \lambda-\frac{N}{(N+1)(N+2)}g\right)\textrm{Tr}(h) \crcr
    & +g\sum_{\alpha}\sum_{k}(e_N)_i^{\alpha}(e_N)_i^{\alpha}(e_N)_k^{\alpha}(e_N)_k^{\alpha}h_k \, ,
\end{align}

\begin{align}
    \beta_{ij}=& g\sum_{\alpha}\sum_{k}(e_N)_i^{\alpha}(e_N)_j^{\alpha}(e_N)_k^{\alpha}(e_N)_k^{\alpha}h_k \quad , \quad 2\leq i <j \, . \label{eq:tetraoffdiag} 
\end{align}

It is possible to use the recursive definition of the $e_N$ vectors to rewrite the beta functions more explicitly. Defining $C_{N,ij}(h)=\sum_{\alpha}\sum_{k}(e_N)_i^{\alpha}(e_N)_j^{\alpha}(e_N)_k^{\alpha}(e_N)_k^{\alpha}h_k$, we have
\begin{equation}
    C_{N,ij}(h)=\frac{1}{\sqrt{i(i+1)j(j+1)}}\left[(1-i) h_i+\sum_{k=1}^{i-1}h_k\right]+\delta_{ij}\left[\frac{i}{i+1}h_i+\sum_{k=i+1}^{N}\frac{h_k}{k(k+1)} \right] \, .
    \label{eq:tetraexpl}
\end{equation}

Solving \eqref{eq:tetraoffdiag} gives $h_1=h_2=\dots h_{N-1}$. We can then simplify the other beta functions, and the fixed-point system reduces to the following two equations:

\begin{align}
    \beta_1&=h_1^2+\left( (N+1)\lambda +\frac{N-2}{N(N+2)}g-\epsilon\right) h_1 + \left(\lambda -\frac{N-2}{N(N+2)}g\right)h_N \, , \crcr
    \beta_N&=h_N^2+\left( 3\lambda +\frac{(N-1)(N-2)}{N(N+2)}g-\epsilon\right)h_N+\left((N-1)\lambda -\frac{(N-1)(N-2)}{N(N+2)}g \right)h_1 \, . 
\end{align}

There are two bulk fixed points with hypertetrahedral symmetry, referred to as $T_{N-}$ and $T_{N+}$, and given by
\begin{equation}
    \lambda_-=\frac{2(N+1)}{3(N+2)(N+3)}\epsilon \; , \; g_-=\frac{N+1}{3(N+3)} \epsilon\, ,
\end{equation}
and 
\begin{equation}
    \lambda_+=\frac{(N-1)(N-2)}{3(N+2)(N^2-5N+8)} \epsilon\; , \; g_+=\frac{(N-4)(N+1)}{3(N^2-5N+8)} \epsilon\, .
\end{equation}
At the $T_{N-}$ fixed point we then have
\begin{equation}
    \beta_{1-}=h_{1-}^2 - \epsilon\frac{N(N+8)+1}{3N(N+3)}h_{1-}+\epsilon\frac{N+1}{3N(N+3)}h_{N-} \, ,
\end{equation}
\begin{equation}
    \beta_{N-}=h_{N-}^2-\epsilon\frac{2N^2+7N-1}{3N(N+3)}h_{N-}+\epsilon\frac{(N+1)(N-1)}{3N(N+3)}h_{1-} \, .
\end{equation}

Besides the trivial fixed point, we have three fixed points given by:
\begin{equation}
h_{1-}=h_{N-}=\frac{N+7}{3(N+3)}\epsilon \, ,  
\end{equation}
\begin{align}
    h_{1-}&=\frac{N(N+9)+2\pm \sqrt{N^3(N+10)+45N^2+4N+4}}{6N(N+3)} \, , \crcr h_{N-}&=\frac{N(3N+7)-2\mp \sqrt{N^3(N+10)+45N^2+4N+4}}{6N(N+3)} \, .
    \label{eq:nttm}
\end{align}

The first fixed point does not break the bulk symmetry, while the other two fixed points break it to $T_{N-1}\times \mathbb{Z}_2$.

And at the $T_{N+}$ fixed point we have
\begin{equation}
    \beta_{1+}=h_{1+}^2 - 2\epsilon\frac{N^2(N-6)+12N-2}{3N(N^2-5N+8)}h_{1+}+\epsilon\frac{2(N-2)}{3N(N^2-5N+8)}h_{N+} \, ,
\end{equation}
\begin{equation}
    \beta_{N+}=h_{N+}^2-\epsilon\frac{2N^2(N-5)+16N+4}{3N(N^2-5N+8)}h_{N+}+\epsilon\frac{2(N-1)(N-2)}{3N(N^2-5N+8)}h_{1+} \, .
\end{equation}

Besides the trivial fixed point, we have three fixed points given by:
\begin{equation}
h_{1+}=h_{N+}=\frac{2N(N-6)+22}{3(N^2-5N+8)}\epsilon \, ,  
\end{equation}
\begin{align}
    h_{1+}&=\frac{N^2(N-6)+13N-4+2\pm \sqrt{N^6-12N^5+58N^4-136N^3+141N^2-32N+16}}{3N(N^2-5N+8)} \, , \crcr h_{N+}&=\frac{N^2(N-4)+5N+4\mp \sqrt{N^6-12N^5+58N^4-136N^3+141N^2-32N+16}}{3N(N^2-5N+8)} \, .
    \label{eq:nttp}
\end{align}

Again, the first fixed point does not break the bulk symmetry while the other two fixed points break it to $T_{N-1}\times \mathbb{Z}_2$.

We can then study stability for some small values of $N$. For the $T_{N-}$ fixed point, we find that for $4\leq N \leq 10$, the trivial defect is unstable and the fixed points of \eqref{eq:nttm} have at least one unstable direction. The fixed point with all defect couplings equal is stable for $4\leq N \leq 6$, has eight stable directions and twenty marginal directions for $N=7$, and is fully unstable for $8\leq N \leq 10$.

For the  $T_{N+}$ fixed point, we find that for $4\leq N \leq 10$, the trivial defect is unstable and the fixed points of \eqref{eq:nttm} have at least one unstable direction while the fixed point with all defect couplings equal is stable. 

Similarly to what we found for the other bulks, the only stable defect fixed point is the one that does not break the bulk symmetry.

\subsection{$O(m)\times O(n)$ biconical model}

The $O(m)\times O(n)$ biconical model has $m$ fields transforming under $O(m)$ and $n$ fields transforming under $O(n)$. It is given by the action
\begin{equation}
    S=\int d^d x \left( \frac{1}{2} \partial_{\mu} \phi_i \partial^{\mu} \phi_i +\frac{1}{2} \partial_{\mu} \psi_j \partial^{\mu} \psi_j + \frac{\lambda_1}{8}\left(\phi ^2\right)^2 + \frac{\lambda_2}{8}\left(\psi ^2\right)^2 + \frac{g}{4}\phi^2 \psi^2 \right) \, ,
\end{equation}
where $\phi_i \, , \, i=1,\dots , m$ are the $m$ fields transforming under $O(m)$ and  $\psi_j \, , \, j=1,\dots , n$ are the $n$ fields transforming under $O(n)$. 

This model does not fall into the class of models described by \eqref{eq:action} and \eqref{eq:decomp} but can be treated in a similar manner \cite{Osborn:2017ucf}. 
Fixed points have been computed for generic values of $m$ and $n$, but have complicated expressions. In the case $m=n$, they are given by
\begin{equation}
    \lambda_1=\lambda_2=\frac{n}{2(n^2+8)}\epsilon \quad  , \quad g=-\frac{n-4}{2(n^2+8)} \epsilon \;  .
\end{equation}

We can then add a surface defect to the action
\begin{equation}
    S_{\text{defect}}=\int dx_1 dx_2 \frac{h_{1,i}}{2}\phi_i^2 +\int dx_1 dx_2 \frac{h_{2,j}}{2}\psi_j^2 \, .
\end{equation}
The beta functions are computed similarly to section \ref{sec:model}, and we obtain
\begin{align}
    \beta_{1,i}&= -\epsilon h_{1,i} +h_{1,i}^2 +\lambda_1\left( \mathrm{Tr}(h_1)+2h_{1,i}\right) +g \mathrm{Tr}(h_2) \, , \crcr
    \beta_{2,j}&= -\epsilon h_{2,j} +h_{2,j}^2 +\lambda_2\left( \mathrm{Tr}(h_2)+2h_{2,j}\right) +g \mathrm{Tr}(h_1)  \, .
\end{align}
Similarly to what happened for the $O(N)$ model and the hypercubic model, the couplings $h_{1,i}$ all respect the same quadratic equation, and all $h_{2,i}$ couplings respect the same quadratic equation. 
Therefore these equations are solved by taking  $l$ of the $h_{1,i}$ couplings equal to $h_{1,l}$ and $m-l$ equal to $ h_{1,m-l}$ and taking $k$ of the $h_{2,j}$ couplings equal to $h_{2,k}$ and $n-k$ equal to $ h_{2,n-k}$, with $0 \leq l \leq m$ and  $0 \leq k \leq n$. 
It means that the original $O(m)\times O(n)$ symmetry will be broken to $\left(O(l)\times O(m-l)\right)\times\left(O(k)\times O(n-k)\right)$.

At first order in $\epsilon$, the system of equations becomes
\begin{align}
    0&= h_{1,l}^2 +h_{1,l}\left( (l+2)\lambda_1-\epsilon\right) +\lambda_1 (m-l)h_{1,m-l}+g \left( k h_{2,k}+ (n-k)h_{2,n-k}\right) \, , \crcr
   0&= h_{1,m-l}^2 +h_{1,m-l}\left( (m-l+2)\lambda_1-\epsilon\right) +\lambda_1 l h_{1,l}+g \left( k h_{2,k}+ (n-k)h_{2,n-k}\right) \, , \crcr
   0&= h_{2,k}^2 +h_{2,k}\left( (k+2)\lambda_2-\epsilon\right) +\lambda_2 (n-k)h_{2,n-k}+g \left( l h_{2,l}+ (m-l)h_{2,m-l}\right) \, , \crcr
   0&= h_{2,n-k}^2 +h_{2,n-k}\left( (n-k+2)\lambda_2-\epsilon\right) +\lambda_2 k h_{2,k}+g \left( l h_{1,l}+ (m-l)h_{1,m-l}\right) \, .
\end{align}

It is possible to solve these equations in all generality, but the fixed points are complicated functions of $m$ and $n$, which we will not give here.

Let us look instead at a simpler case: $n=m$ and only two surface defect couplings $h_{1,i}=h_1$ for all $i=1,\dots m$, $h_{2,j}=h_2$ for all $j=1,\dots n$. The beta functions become
\begin{align}
    \beta_1 &= h_1^2-\frac{n^2-2n+16}{2(n^2+8)}\epsilon h_1 -\frac{n(n-4)}{2(n^2+8)}\epsilon h_2 \, , \crcr
    \beta_2 &= h_2^2-\frac{n^2-2n+16}{2(n^2+8)}\epsilon h_2 -\frac{n(n-4)}{2(n^2+8)}\epsilon h_1 \, .
\end{align}

There are three solutions besides the trivial one given by
\begin{equation}
    h_{1,1}=h_{2,1}=\frac{n^2-3n+8}{n^2+8}\epsilon \, ,
    \label{eq:bicoeq}
\end{equation}
and
\begin{equation}
    h_{1,\pm}=\frac{n+8\pm \sqrt{2n^3+9n^2-48n+64}}{2(n^2+8)}\epsilon \, , \,  h_{2,\pm}=\frac{n+8\mp \sqrt{2n^3+9n^2-48n+64}}{2(n^2+8)}\epsilon \, .
    \label{eq:bicodif}
\end{equation}

We can then compute the stability matrix and critical exponents. For the trivial fixed point the critical exponents are $-\frac{n+8}{n^2+8}\epsilon$ and $-\frac{n^2-3n+8}{n^2+8}\epsilon$ which are both negative for all values of $n$. Therefore, the trivial defect fixed point is unstable. 
For the solution of \eqref{eq:bicoeq}, the critical exponents are given by
\begin{equation}
    \Vec{\omega}_1=\left(\frac{n^2-3n+8}{n^2+8} \epsilon\, , \, \frac{2n^2-7n+8}{n^2+8} \epsilon\right) \, ,
\end{equation}
which are both positive. Therefore, the fixed point \eqref{eq:bicoeq} is stable. 
For the two solutions of \eqref{eq:bicodif}, the critical exponents are given by
\begin{equation}
    \omega_{\pm}=\frac{-n(n-4)\pm \sqrt{n^4+52n^2-192n+256}}{2(n^2+8)} \epsilon\, .
\end{equation}
$\omega_+$ is positive for all values of $n$ while $\omega_-$ is negative for all values of $n$. Therefore, the two fixed points of \eqref{eq:bicodif} always have one stable direction and one unstable direction.

\section{Conclusion}

While line defects and interfaces have been studied for generic critical multiscalar models in \cite{Pannell:2023pwz} and \cite{Harribey:2024gjn}, respectively, surface defects had only been considered for the $O(N)$ model \cite{Giombi:2023dqs, Trepanier:2023tvb, Raviv-Moshe:2023yvq}, a free bulk \cite{Bashmakov:2024suh} or for symmetry-preserving defects in \cite{Pannell:2024hbu}. In this paper, we filled this gap and studied generic surface defects for multiscalar models. 

This study can be done in a more systematic way than for line defects and interfaces. Indeed, as the surface defect is given by a symmetric matrix $h_{ij}$, fixed points are characterised only by their eigenvalues. We were thus able to restrict our study to a diagonal defect matrix, which greatly simplified the computations. 

Moreover, the remaining symmetry at the defect fixed point depends only on the number of different eigenvalues of $h_{ij}$ and their multiplicity. In general, for a bulk symmetry $G_N$ an eigenvalue of multiplicity $1<l<N$ will give a smaller copy $G_l$ of the bulk symmetry while an eigenvalue of multiplicity one will just give a $\mathbb{Z}_2$ symmetry (the action is always symmetric when flipping the sign of a single field). 
More precisely, in the case of the $O(N)$ model and the hypercubic model, we found fixed points where $h_{ij}$ had an eigenvalue of multiplicity $l$ and an eigenvalue of multiplicity $N-l$ for $0\leq l \leq N$ breaking the bulk symmetries to $O(l)\times O(N-l)$ and $B_l \times B_{N-l}$ respectively. For the hypertetrahedral bulk, the surface defect matrix $h_{ij}$ had at the fixed point one eigenvalue of multiplicity $N-1$ and one eigenvalue of multiplicity one, giving a $T_{N-1}\times \mathbb{Z}_2$ symmetry. 
We expect other bulk critical models to undergo similar patterns of symmetry breaking. However, due to the quadratic nature of the surface defect, there will always be at least a $\mathbb{Z}_2^N$ symmetry. 

We also studied the stability of defect fixed points, and for all the bulks we considered, the only stable defect was the one with one non-zero eigenvalue of multiplicity $N$, which does not break the bulk symmetry. 

Our results show that going beyond the $O(N)$ model, surface defects can lead to a greater variety of symmetry breakings, even though these are more constrained than in the case of interfaces. Of course, these results are perturbative in $\epsilon$, and pushing the computations to higher loops could uncover new patterns of symmetry breaking. Analytic bootstrap methods could also be used to study the spectrum and operator product expansions of surface defects CFTs \cite{Dey:2020jlc}. For example, generalising the methods of \cite{Gliozzi:2015qsa,Behan:2020nsf, Behan:2021tcn} to surface defects would be an interesting extension of our work. Another possible extension of our work would be to study multiple surface defects and the resulting effective field theory, as recently discussed for boundaries in \cite{Diatlyk:2024qpr, Kravchuk:2024qoh}.

\section{Acknowledgments}

We would like to thank Andreas Stergiou and William H. Pannell for useful discussions and comments on an early draft of the paper. AA thanks the organizers of the Nordita Summer Internship Program, during which part of this research was carried out.


\addcontentsline{toc}{section}{References}

\providecommand{\href}[2]{#2}\begingroup\raggedright\endgroup

\end{document}